
\input harvmac.tex


\let\d\partial

\def\frac#1#2{{\textstyle{#1\over #2}}}


\def\text#1{\quad\hbox{#1}\quad}

\def\la{\lambda}
\def\e{\epsilon}
\def\ta{\theta}
\def\a{\alpha}

\let\Rw\Rightarrow

\def\rw{\rightarrow}





\newcount\eqnum  
\eqnum=0
\def\eq{\eqno(\secsym\the\meqno)\global\advance\meqno by1}
\def\eqlabel#1{{\xdef#1{\secsym\the\meqno}}\eq }  

\newwrite\refs 
\def\startreferences{
 \immediate\openout\refs=references
 \immediate\write\refs{\baselineskip=14pt \parindent=16pt \parskip=2pt}
}
\startreferences

\refno=0
\def\aref#1{\global\advance\refno by1
 \immediate\write\refs{\noexpand\item{\the\refno.}#1\hfil\par}}
\def\ref#1{\aref{#1}\the\refno}
\def\refname#1{\xdef#1{\the\refno}}

\newcount\exno
\exno=0
\def\Ex{\global\advance\exno by1{\noindent\sl Example \the\exno:

\nobreak\par\nobreak}}

\parskip=6pt

\overfullrule=0mm

\def\frac#1#2{{#1 \over #2}}		

\let\ta=\theta
\let\d=\partial
\let\La=\Lambda

\def\rw{{\rightarrow}}

\newwrite\refs
\def\startreferences{
 \immediate\openout\refs=references
 \immediate\write\refs{\baselineskip=14pt \parindent=16pt \parskip=2pt}
}
\startreferences

\refno=0
\def\aref#1{\global\advance\refno by1
 \immediate\write\refs{\noexpand\item{\the\refno.}#1\hfil\par}}
\def\ref#1{\aref{#1}\the\refno}
\def\refname#1{\xdef#1{\the\refno}}


\Title{\vbox{\baselineskip12pt
\hbox{LAVAL-PHY-{99-21}}}}
{\vbox {\centerline{ Open problems for the superKdV equations }}}

\smallskip
\centerline{ P. Mathieu\foot{Work
supported by NSERC (Canada) and FCAR (Qu\'ebec) }\foot{Talk 
presented at AARMS-CRM
Workshop on Backlund and Darboux Transformations:  The geometry of Soliton Theory.
June 4-9, 1999 (Halifax, Nova Scotia).} }

\smallskip\centerline{ \it D\'epartement de
Physique,} \smallskip\centerline{Universit\'e Laval,}
\smallskip\centerline{ Qu\'ebec, Canada G1K 7P4}
\smallskip\centerline{pmathieu@phy.ulaval.ca}
\vskip .2in
\bigskip

\bigskip
\centerline{\bf Abstract}
\bigskip
\noindent
 After a review of the 
basic results concerning the $N=1,2$
supersymmetric extensions of the Korteweg-de Vries equation, with a pedagogical
presentation of the superspace techniques, we discuss some basic open problems mainly
in  relation with the
$N=2$ extensions.

\leftskip=0cm \rightskip=0cm

\Date{07/99\ } 

\let\n\noindent


\newsec{Introduction}

Supersymmetry offers a powerful tool for widening the scope of integrability.  The
field of supersymmetric integrable systems turns out to be remarkably rich in
addition to further displaying novel features such as conserved nonlocal `Poisson
square roots' of local conservation laws.\foot{Some physical motivations for
considering supersymmetric integrable systems are scattered in the footnotes 3, 4
and 7, while geometrical implications are alluded to in the conclusion.} Not
surprisingly, it started with the extension of the 
Korteweg-de Vries (KdV) equation although by now many other equations have been
supersymmetrized.

The $N=1,2$ (where $N$ refers to the number of supersymmetries)  integrable
supersymmetric versions of the KdV equation have been found about
10 years ago [\ref{P. Mathieu, {\it Supersymetric extension of the
Korteweg-de Vries equation}, J. Math. Phys. {\bf 29} (1988) 2499-2506; {\it
Superconformal algebra and supersymmetric Korteweg-de Vries equation}, Phys. Lett.
{\bf B203} (1988) 287-291.}\refname\Mat, \ref{Yu. I. Manin and A.O. Radul, {\it A
supersymmetric extension of the Kadomstev-Petviashvili hierarchy}, Comm. Math, Phys.
{\bf 98} (1985) 65-77.}\refname\MR,
\ref{C.A. Laberge and P. Mathieu, {\it $N=2$ superconformal algebra and $O(2)$
integrable fermionic extensions of the Korteweg-de Vries equation}, Phys. Lett.
{\bf B215} (1988) 718-722;  P. Labelle and P. Mathieu, {\it  A new N = 2
supersymmetric Korteweg-de Vries equation}, J. Math. Phys. {\bf 32} (1991)
923-927.}\refname\La,
\ref{ Z. Popowicz, {\it The Lax formulation of the `new' $N=2$ SUSY KdV equation},
Phys. Lett. {\bf 174} (1993) 411.}\refname\Po].\foot{The present restriction to
$N\leq 2$ is partly motivated by the limitations of my own works. However there is
also a physical motivation:  these systems have found a remarquable application,
albeit in their quantum formulation, in perturbed conformal field
theory [\ref{B.A. Kupershmidt and P. Mathieu, {\it Quantum
Korteveg-de Vries like equations and perturbed conformal field theories}, Phys.
Lett. {\bf B227} (1989) 245-250;  P. Mathieu, {\it Integrability of perturbed
superconformal minimal models}, Nucl. Phys. {\bf B336} (1990) 338-348; 
P. Mathieu and M.A. Walton,  {\it Integrable perturbations
of N = 2 superconformal minimal models}, Phys. Lett. {\bf  B254} (1991) 106-110.}], in the context of which 
$N=2$ is the maximal number of supersymmetries that is of real interest.
Nevertheless, in some context, the $N=3$ and $4$ extensions may be physically relevant
and the corresponding extensions of KdV have been considered; see for instance
[\ref{C.M. Yung, {\it The $N=3$ supersymmetric KdV hierarchy}, (1993); S. Krivonos, A.
Pashnev and Z. Popowicz,{\it  Lax Pairs for
$N=2,3$ SUSY KdV equations and their extensions}, Mod.  Phys. Lett. {\bf A13} (1998)
1435; F. Delduc, L. Gallot and E. Ivanov, {\it New super KdV system with $N=4$ SCA as
the hamiltonian structure},  Phys.Lett. {\bf B 396} (1997) 122-132.}].
Our discussion will also be restricted to supersymmetric KdV extensions having an even
Poisson brackets.  A new $N=2$ super KdV equation with an odd Poisson structure has
been found recently in [\ref{Z. Popowicz,
  {\it Odd bihamiltonian structures of new
 supersymmetric $N= 2,4$ Korteweg-de Vries equation and odd susy Virasoro-like
 algebra}, 
Phys. Lett {\bf B459} (1999) 150-158.}] but the resulting equations are somewhat less
interesting in that the bosonic fields do not couple to the fermions in their time
evolution.  It would be of interest to see whether there are other `odd' integrable
extensions displaying fermionic interactions in the bosonic evolution equations.}  
The key points of this development were 1- the realization that supersymmetrization
could be restricted to the space variable only and 2- that the  crucial KdV struture
whose core needs to be preserved is the KdV second hamiltonian structure.\foot{Here
are some comments on the literature concerning the
$N=1$ case. Already in 1984, Kupershmidt [\ref{B. Kupershmidt,  {\it
A super Korteweg-de Vries equation: An integrable system}, Phys. Lett. {\bf A 102}
(1984) 213-215.}\refname\Ku ]  has presented a simple fermionic (but not
supersymmetric - see below) extension of the KdV equation by placing the emphasis on
its hamiltonian formulation (this system has actually two local hamiltonian
structures).  This was an important step toward the formulation of the right
supersymmetric extension given  that both hamiltonian operators were supersymmetric
invariant (the hamiltonian themselves were not) and could then serve in the
formulation of a genuine supersymmetric system. The supersymmetric KdV equation was
initially  found independently of the work of Manin-Radul [\MR] on super KP
hierarchy.  
It was realized afterwards that this general system has
indeed a reduction to the supersymmetric KdV system. 
} In Fourier components, the underlying hamiltonian
operator is the Poisson bracket formulation of the Virasoro algebra [\ref{J.L.
Gervais, {\it Infinite familiy of plynomial functions of teh Virasoro
generators with vanishing Poisson brackets}, Phys. Lett. {\bf B160} (1985) 277-279;
B.A. Kupershmidt,  {\it Super Korteweg-de Vries equations associated to
super extensions of the Virasoro algebra}, Phys. Lett. {\bf A109} (1985) 417-423.}],
for which supersymmetric extensions were already known  and could then be used to
construct the appropriate supersymmetric extension of KdV.  A number of developments
have occured in the following years but recently the focus has  moved toward the
construction of extended 
 super KdV hierachies\foot{The most important physical application of these
constructions concerns conformal field theory: the corresponding Poisson structures
yield clasical versions of super $W$ algebras whose quantum form can be obtained
via the quantization of the modified fields obtained through the Miura
transformation (see e.g. [\ref{P. Mathieu, {\it Extended classical conformal algebras
and the second Hamiltonian structure of Lax equations}, Phys. Lett. {\bf B208}
(1988) 101-106.}]).} (in
the way the Boussinesq equation generalizes  KdV, that is, via higher order Lax
operators) [\ref{C.M. Yung, {\it The $N=2$ supersymmetric Boussinesq
hierarchies}, Phys. Lett. {\bf B309} (1993) 75-84; Z. Popowicz, {\it Lax formulation of
the $N=2$ superBoussinesq equation}, Phys. Lett. {\bf B319} (1993) 478-484; 
 {\it Extensions of the $N=2$ SUSY $a=-2$ Boussinesq hierarchy}, Phys.Lett {\bf A 236}
(1997) 455-461;  S. Belluci, E. Ivanov,
S. Krivonos and A. Pichugin, {\it $N=2$ super Boussinesq hierarchy: Lax pairs and
conservation laws}, Phys. Lett. {\bf B312} (1993) 463-470;  F.Delduc and L.Gallot, {\it
$N=2$ KP and KdV hierarchies in extended superspace}, Comm. Math. Phys. {\bf 190}
(1997) 395-410.}]. However a certain number of problems associated to the
$N=1,2$ KdV systems have remained unsolved and the goal of this presentation is to
identify some of them.  For this it is necessary to present a brief review of the
supersymmetric formulation of the KdV equation and, for the benefit of those readers
unfamiliar with super technologies, some manipulations will be worked out in some
detail.

\newsec{Supersymmetrization of the KdV equation: N=1}

We will formulate the supersymmetric extension of the KdV equation in the
{\it superspace formalism}.  That amounts to extend the $x$ variable to a doublet
$(x,\ta)$ where
$\ta$ is a Grassmannian (or {\it anticommuting}) variable: $\ta^2 =0$.  Ordinary
(i.e. commuting) fields
$f(x)$  (functions of $x$ and $t$ in fact but the time dependence will generally be
suppressed) will be replaced by {\it superfields} $F(x,\ta)$. Given that
$\ta^2=0$, these superfields have a very simple Taylor expansion in terms of
$\ta$:
$$F(x,\ta) = f(x)+\ta \gamma(x)\eq$$
$f$ and $\gamma$ are called the {\it component fields}.  $\gamma$ is said to be the
super-partner of $f$ and vice-versa.  In the present case, $F(x,\ta)$ is a {\it
bosonic} superfield: it has the same `statistics' (i.e. commuting or
anticommuting character) as the field appearing in the
$\ta$ independent term (here
$f$); on the other hand,
$\gamma$ is anticommuting, i.e. it is a {\it fermionic field}.  In particular,
$\gamma(x)\gamma(y) = - \gamma(y)\gamma(x)$ so that $\gamma(x)^2=0$; also for
instance, $\ta\gamma= -\gamma\ta$.  The final ingredient that we need is the {\it
superderivative} $$D= \ta\d+\d_\ta\eq$$ whose square is the usual space derivative:
$D^2=\d$.

A supersymmetry transformation is nothing but a translation in superspace. Such a
translation takes the form:
$x\,\rw\, x-\eta\ta $ and $\ta\,\rw\, \ta+\eta$, where $\eta$ is a constant
anticommuting parameter, supposed, in the following, to be very small.  Consider
then the effect of the translation in the superfield: 
$$ \eqalign{  F (x, \ta) \rw F (x-\eta\ta, \ta+\eta)  &= F (x, \ta) -\eta \ta \d
F(x,\ta) + \eta \d_\ta F(x,\ta)\cr
& \equiv F (x, \ta)+
\delta_\eta F(x,\ta)\cr&= f(x)+\ta \gamma (x) + \delta_\eta f(x) +
\ta \delta_\eta \gamma(x)\cr}
 \eq$$ 
The second equality shows that $\delta_\eta$ is bosonic so
that it commutes with
$\ta$. 
We read off the component-field transformations to be
$$ \delta_\eta f = \eta \gamma \quad, \qquad \delta_\eta \gamma =\eta
f_x \eqlabel\transa$$ This is called a {\it supersymmetry transformation}; it
relates a bosonic field to a fermionic field and vice-versa. It has the
remarquable virtue of linking a field transformation to a (super)space
translation.  Note that two successive supersymmetry transformations lead to
$$ \delta_\eta\delta_{\eta'} \, f = \eta' \eta\, f_x \quad,\qquad\quad \delta_\eta
\delta_{\eta'}\, \gamma =\eta'\eta \, \gamma_x\eq$$  
In other words, a translation in superspace, hence a
supersymmetry transformation, is a sort of square root of an ordinary translation.
Every local expression in the
superfields and the superderivatives is manifestly supersymmetric invariant.

To supersymmetrize the KdV equation
$$ u_t= -u_{xxx}+6uu_x\eq$$ one should then start by extending the $u$
field to a superfield.  There are two ways of doing this: either as a fermonic
superfield
$$u(x)\quad \rw \quad \phi(x,\ta) = \ta u(x)+\xi(x)\eq$$
or as a bosonic superfield
$$u(x)\quad \rw \quad  U(x,\ta) = u(x)+\ta\la(x)\eq$$
 It turns out that  the first choice is the one that gives the
interesting extension.\foot{The other possibility would not have a second
hamiltonian structure associated to the super Virasoro algebra as a simple
dimensional (i.e. degree counting) analysis shows (it requires the introduction
of an anticommuting field of degree $3/2$ as $\xi$ and not of degree $5/2$ as
$\la$). The degree counting is explained below. Notice that to a large extend, we try
to reserve Greek letters for anticommuting variables or fields.}
The KdV equation is homogeneous with respect to the scaling gradation: in the
normalization where deg $\d= 1$, one finds that deg $u=2$.  The identity $D^2=\d$
implies that  deg
$D=1/2$, so that deg
$\ta=-1/2$. For the superfield to be homogeneous, $\xi$ must have degree $3/2$.

Let us then proceed with a direct extension of the KdV equation, multiplying
each term by $\ta$ and rewriting the result in terms of superfields :
$$\eqalign{
 u_t &\;\rw\; \, \phi_t\cr
 u_{xxx}&\;\rw\; \, \phi_{xxx}\cr
3uu_x &\;\rw\; \, c \phi D\phi_x+(6-c)\phi_x(D\phi)\cr}\eq$$
where $c$ is a free constant. 
We thus observe that the nonlinear term  does not have a unique extension in
terms of superfields.  Therefore, this  direct extension leaves us  with a
supersymmetric version of the KdV equation containing a free parameter:
$$\phi_t= \phi_{xxx}+c (\phi D\phi)_x+(6-2c)\phi_x(D\phi)\eqlabel\skdv$$ It
turns out that this equation is integrable only if $c=3$ [\Mat]\foot{Actualy the case
$c=0$ is also integrable but its leads to a somewhat trivial system in which the
fermionic fields decouple from the bosonic equation which reduces then to the usual
KdV equation.  Nevertheless, this equation happens to be relevant in supersymmetric
extensions of matrix models that describes superstrings in $d<3/2$ dimensions, or
equivalently, conformal field theories coupled to gravity [\ref{K. Becker and M.
Becker, {\it Nonperturbative solution of the super-Virasoro constraints}, Mod. Phy.
Lett. {\bf A8} (1993) 1205-1214.}].}.  We call the resulting equation the {\it super
KdV equation}, or sKdV for short.  Its component version reads\foot{Notice that the
product
$\xi\xi_{xx}$  acts as a bosonic field: quite
generally, a product of two fermions is a boson.  It can be seen 
easily that  passing an anticommuting variable in front of it does
not induce an overall minus sign, e.g. $\xi\xi_{xx}\ta= -\xi\ta\xi_{xx}=
\ta\xi\xi_{xx}$. Notice moreover that this term is a total derivative:
$\xi \xi_{xx}= (\xi\xi_x)_x$ since the extra  resulting term is
$\xi_x\xi_x=0$.}
$$\eqalign{
&u_t= -u_{xxx}+6uu_x- 3\xi\xi_{xx}\cr
&\xi_t= -\xi_{xxx}+3(u\xi)_x\cr}\eqlabel\skdvco$$

It is not difficult to verify that the  system (\skdvco) is invariant under
the supersymmetry transformation $ \delta_\eta u = \eta \xi_x $ and $ \delta_\eta \xi
=\eta u$. 
This is not so for the integrable fermionic extension proposed by
Kupershmidt [\Ku]:
$$\eqalign{
&u_t= -u_{xxx}+6uu_x- 3\xi\xi_{xx}\cr
&\xi_t= -4\xi_{xxx}+6u\xi_x +3u_x\xi\cr}\eqlabel\kuper$$
With (\skdvco) being called the super KdV equation, it would be appropriate to
call (\kuper) the {\it Kuper-KdV equation}.

The integrability of (\skdv) can be established in various ways. The most
direct argument is that it has a Lax representation:\foot{Actualy, the Lax operator
is not unique: the choice $L= \d^2+\phi D-(D\phi)$ (the formal adjoint of
$\d^2-\phi D$) leads to completely equivalent results.}
$$L_t= [-4L^{3/2}_+,L]\qquad L=\d^2-\phi D\eq$$ 
and the conservation laws are obtained as follows (the subscript gives the degree):
$$H_{2k+1}= \int dx d\ta \; {\rm sRes} L^{(2k+1)/ 2}\eq$$
For super pseudodifferential operators, the $+$ projection and the super
residue sRes are defined as follows
$$\Lambda= \sum_{k=-\infty}^N \a_i D^i, \qquad \Lambda_+= 
\sum_{k=0}^N \a_i D^i\qquad {\rm sRes}\Lambda= \a_{-1}\eq$$
In the above expression for the conservation laws, we have also introduced
the superintegration. The integration over the $\ta$ variable is defined as
follows:
$$\int d\ta \, 1 = 0 \qquad \int d\ta \, \ta = 1\eq$$
The integration over $\ta$ is thus essentially equivalent to the
differentiation with respect to $\ta$. With these rules, the
superintegration of a superderivative vanishes (with the usal rule that the
ordinary  integral of a total derivative vanishes):
$$\int dx d\ta \, [D\phi(x,\ta)] = 
 \int dx d \ta \, (\ta \xi_x+u)= \int dx \, \xi_x=0\eq$$ For instance, the second
conservation law is\foot{Notice that for a fermionic variable, $\xi \xi_x$ is {\it not} a total derivative: a
partial integration of  $\int dx \, \xi\xi_x$ leads to $ -\int dx\, \xi_x\xi$ and the
interchanges of the two terms generates another minus sign so that the original
expression is recovered. }
$$H_3= \int dx d\ta \;(\phi D\phi)= \int dx \;(u^2-\xi\xi_x)\eq$$

Another way of establishing the integrability is to supersymmetrize the Gardner
transformation [\ref{C.S. Gardner, {\it Korteweg-de vries equation and
generalizations. IV. The  Korteweg-de vries equation as a hamiltonian system}, J.
Math Phys. {\bf 19} (1971) 1548-1551.}].  This extension is unique: 
$$\phi= \chi+\e\chi_x+\e^2\chi D\chi\eqlabel\sG$$ with $\chi= \ta w+\sigma$. It maps
a solution of the super Gardner equation\foot{The component form of this superfield
equation reads:
$$\eqalign{
&w_t = -w_{xxx}+6ww_x-\sigma\sigma_{xx} +\e^2 [6w^2w_x -3(\sigma \sigma_x w)_x]\cr
&\sigma_t= -\sigma_{xxx}+3(\sigma w)_x+\e^2[3w(w\sigma)_x]\cr}$$ 
Notice that $\sigma_t$ is not a total derivative. The usual Gardner equation is
recovered by setting the fermionic field
$\sigma=0$ and the sKdV equation is the limiting case where $\e=0$.}
$$\chi_t= -\chi_{xxx}+3(\chi D\chi)_x+\e^2 \, 3(D\chi)(\chi D\chi)_x\eq$$
into a solution of the sKdV equation.  Since
$\chi_t$ is a total superderivative, e.g.
$$(D\chi)(\chi D\chi)_x = \frac16 D[(D\chi)^3]+\frac12[\chi(D\chi)^2]_x\eq$$
$\int dx d\ta\,  \chi$ is conserved and by inverting the super Gardner transformation
(\sG), we recover an infinite number of conservation laws:
$$\chi= \sum_{n=0}^\infty \e^n \, h_n[\phi]\quad \Rw\quad {d\over dt} \int dx d\ta
\sum_{n=0}^\infty \e^n \, h_n[\phi]= 0 \eq$$ (where $h_n[\phi]$ stands for a
differential polynomial in $\phi$).  Now the crucial point is that the sKdV equation
is independent of $\e$ so that each separate power of $\e$ must be separately
conserved.  This produces an infinite number of conservation laws, half of which can
be shown to be nontrivial, having a leading term $\phi(D\phi)^{k}$; these are bound to
be the $H_{2k+1}$ above [\Mat]. Note that these are all bosonic
($\chi$ is fermionic but the measure $dx d\ta$ is also fermionic).

Finally, we point out that the sKdV equation is bihamiltonian, the two hamiltonian
operators being\foot{The second hamiltonain structure has been found in the first two
references and the first one in the last two.  Here and below, the action of the
derivatives is always delimited by parentheses, e.g., $D\phi= (D\phi)-\phi D$.}
[\Mat,\ref{I. Yamanaka and R. Sasaki, {\it Super Virasoro algebra and solvable
supersymmetric quantum field theories},  Prog. Theor. Phys. {\bf 79} (1988)
1167-1184.},\ref{W. Oevel and Z. Popowicz, {\it The bi-hamiltonian structure of
the fully supersymetric  Korteweg-de vries equation}, Comm. Math. Phys. {\bf 139}
(1991) 441-460.}\refname\OP,\ref{J.M. Figueroa-O'Farrill, J. Mas and E. Ramos,
{\it Bihamiltonian structure of the supersymmetric SKdV hierarchy}, Rev. Mod. Phys.
{\bf 3} (1991) 479.}],
$$ P_1= \d [D^3-\phi]^{-1}\d\quad, \qquad  P_2= -D^5+3\phi
\d+(D\phi)D+2\phi_x\eq$$
Notice that $P_1$ is a very complicated nonlocal hamiltonian operator,
being essentially an infinite series: $[D^3-\phi]^{-1}= D^{-3}[1-D^{-3}\phi]^{-1}$. 
$P_2$ is the direct supersymmetrization of the KdV second hamiltonian structure:
$-\d^3+4u\d+2u_x$. 

There is a remarquable feature of the  super case that is not present for the usual
KdV equation which is the presence of {\it fermionic  nonlocal conservation laws}
[\ref{P.H.M. Kersten, {\it Higher order supersymmetries and fermionic
conservation laws of the supersymmetric extension of the KdV and the mKdV
equation}, Phys. Lett. {\bf A134} (1988) 25-30.}, 
\ref{P. Dargis and P. Mathieu, {\it Nonlocal conservation laws for
supersymetric  KdV equations}, Phys. Lett. {\bf A176} (1993) 67-74.}\refname\DM].  The first few of them are
$$\eqalign{ &J_{1/2}= \int dx d\ta\; (D^{-1}\phi)= \int dx \;\xi\cr
&J_{3/2}= \int dx d\ta \;(D^{-1}\phi)^2= \int dx\; u(\d^{-1}\xi)\cr
&J_{5/2}= \int dx d\ta \;[(D^{-1}\phi)^3-6\d^{-1}(\phi D\phi)]= \int dx \;
[3\xi(\d^{-1}u)^2-6\d^{-1}(u^2-\xi\xi_x)]\cr}\eq$$ They Poisson commute with the
local bosonic conservation laws
$H_n$ but not among themselves :
$$\{J_{(4n+i)/2},J_{(4m+i)/2}\}= H_{2(n+m)+i}\quad {\rm with}\;  i=1,3 \qquad
\{J_{(4n+1)/2},J_{(4m+3)/2}\}= 0\eq$$  The $J_i$ are thus some sort of Poisson
square roots of the usual conservation laws. The first fermionic nonlocal
conservation law is actually local in terms of the component fields. It signals the
presence of the supersymmetry invariance (and in particular
$\xi$ is not a conserved density for the Kuper-KdV equation).  The infinite sequence
can be generated from the first two conservation laws by the application of the
recursion operator
$P_1^{-1}P_2$.  But there is a more spectacular way of expressing them that makes
manifest their supersymmetric origin: in superspace, $\d^2$  not only has a square
root but it also has a {\it fourth root}: $(\d^2)^{1/4}= D$.  The fermionic nonlocal
conservation laws are thus related to the super residues of the fourth root of odd
powers of the Lax operators as
$$J_{k/2}= \int dx d\ta\;  {\rm sRes} L^{k/4}\qquad (k\; {\rm odd})\eq$$ The sKdV
equation  represents the first example of an integrable system for which  nonlocal
conservation laws arise in such a clean form. 
In that respect,
we introduce the first open problem ({\bf OP}):

\n {\bf OP-1:} Find an integrable deformation that reproduces the fermionic
nonlocal conservation laws.

\newsec{Supersymmetrization of the KdV equation: N=2}

The next extension to be considered is the addition of an extra supersymmetry which
amounts to add an extra anticommuting space dimension. We thus extend $x$ to a
triplet $(x, \ta_1,\ta_2)$, with $\ta_1^2=\ta_2^2=0, \; \ta_1\ta_2=-\ta_2\ta_1$ and
introduce two super derivatives:
$$D_1= \ta_1\d+\d_{\ta_1}\qquad D_2= \ta_2\d+\d_{\ta_2}, \qquad D_1^2=D_2^2=\d,
\qquad D_1D_2=-D_2D_1\eq$$ The superfields are now functions of $(x,\ta_1,\ta_2)$ (as
well as $t$) and their Taylor expansion in terms of the anticommuting variables contain
four terms. For instance the $N=2$ KdV superfield will be written as:
$$\Phi(x,\ta_1,\ta_2)= \ta_2\ta_1 u(x) + \ta_1\xi_1(x)+\ta_2\xi_2(x)+ v(x)\eq$$
$\xi_1$ and $\xi_2$ are two fermionic fields and $v$ is a new bosonic field of
degree $1$ (in a supersymmetric theory, the number of bosonic and
fermionic fields must be the same). Notice that $\Phi$ is a bosonic superfield.

One could then proceed to the direct supersymmetrization of the KdV equation and get
a multiparameter $N=2$ extension.  However, a sounder approach, that reduces
substantially the number of such free parameters, is to formulate the equation
directly in terms of the $N=2$ supersymmetric version of the second hamiltonian
structure (which is expected to be the core structure underlying integrability of
nontrivial KdV extensions):
$$P_2= D_1D_2\d+2\Phi\d-(D_1\Phi)D_1-(D_2\Phi)D_2+2\Phi_x\eq$$
e.g. as
$$\Phi_t= P_2{\delta \over \delta \Phi}\int dx d\ta_1 d\ta_2 \; [ \Phi (D_1D_2\Phi)+a
\Phi^3]\eq$$
This hamiltonian is  the direct generalisation of the KdV hamiltonian $\int dx \, u^2$
and of the
$N=1$ version $\int dx d \ta \; \phi(D\phi)$.  However, the $N=2$ generalisation is not
unique and this introduces a free parameter $a$. The resulting
equation is
$$\Phi_t= -\Phi_{xxx}+3(\Phi D_1D_2\Phi)_x+{(a-1)\over
2}(D_1D_2\Phi^2)_x+3a\Phi^2\Phi_x\eq$$
This is called the SKdV$_a$ equation (the capital S is used for $N=2$).
This system is integrable for exactly three values of $a$ [\La]: $a=-2,1,4$ . For
$a=-2,4$, the Lax representation is standard: $L_t=[-4L^{3/2}_+,L]$ with
$$\eqalign{
&L_{a=4}= -(D_1D_2+\Phi)^2\cr
&L_{a=-2}= -\d^2+\sum_{i,j=1,2}\e_{ij}D_i(D_1D_2+\Phi)D_j\cr}\eq$$ (with
$\e_{12}=-\e_{21}=1$) while for $a=1$ it is nonstandard [\Po]: $L_t=[-4L^{3}_{\geq
1},L]$ with
$$L_{a=1}= \d-\d^{-1}[(D_1D_2\Phi)-(D_2\Phi)D_1-(D_1\phi)D_2+\Phi D_1D_2]\eq$$
In all cases, there is an infinite number of conservation laws given by
$$H_{2k+1}= \int d d\ta_1 d\ta_2 \; {\rm SRes}L^{(2k+1)/2}\eq$$
where the $N=2$ version of the residue of a pseudodifferential operator is the
coefficient of $D_1D_2\d^{-1}$.

Although there exists a Miura transformation, there are no known integrable (i.e.,
Gardner-type) deformations of it. This leads us to:

\n {\bf OP-2:} Find an integrable deformation for SKdV$_{-2,1,4}$ that reproduces their
conservations laws.

For $a=-2,4$, there are two independent towers of fermionic nonlocal conservations
laws  [\DM]: in each case, the first one is $$\int dx d\ta_1d\ta_2 \;(D_i^{-1}\Phi)=
\int dx
\; \xi_i \qquad (i=1,2)\eqlabel\nlf$$  However, although the Lax operator has two
distinct fourth roots, these have not yet  been related to these
fermionic conservation laws:

\n {\bf OP-3:} For SKdV$_{-2,4}$, find the relation between the fermionic nonlocal
conservation laws and the Lax operator.

The existence of these fermionic conservation laws
is natural in that there are two supersymmetries.  On the other hand, for the $a=1$
case, the first fermionic laws (\nlf) do not generate infinite towers: they are
isolated.

\n {\bf OP-4:} Why there are no infinite towers of fermionic conservation laws for
SKdV$_{1}$?

The SKdV$_{-2,4}$ equations are both bihamiltonian: their first hamiltonian operator
is [\OP]
$$P_1^{(a=4)}= \d \quad, \qquad  P_1^{(a=-2)}= (D_1D_2\d^{-1} - D_1^{-1} \Phi D_1^{-1}-
D_2^{-1} \Phi D_2^{-1}) D_1^{-1} \Phi D_1^{-1}
\eq$$
In that respect, the SKdV$_1$ stands as one of the rare example of classical 
integrable system which is not (known to be) bihamiltonian.

\n {\bf OP-5:} Is SKdV$_{1}$ bihamiltonian?

\n Another very natural question is:

\n {\bf OP-6:} Why is there exactly three integrable $N=2$ super KdV
extensions?\foot{There are in fact also three distinct SKdV hiearchies but they
generalize (in the Lax sens) the SKdV$_{2,4}$ equations; the SKdV$_1$ equation appears
as a sort of isolated point. Interesting technical observations in relation
with the bosonic truncation of the Lax operators are presented in [\ref{L. Bonora, S.
Krivonos and A. Sorin, {\it Toward the construction of $N=2$ supersymmetric
iontegrable hierarchies}, Nucl. Phys. {\bf B477} (1996) 835-854.}].}  Is there an
underlying Lie algebraic  interpretation for this threefold way (i.e, is this
related to the existence of the three classical algebras)? 

\newsec{Concluding questions}

There are further general questions that could be formulated in relation with super
integrable systems. As stressed here, these equations are naturally formulated in
superspace. Over the years, it became clear that a lot of structure is contained in
the Painlev\'e test: for instance its truncation leads to Backl\"und transformations
and the Lax operator [\ref{J. Weiss, M. Tabor and G. Carnavale, {\it The Painlev\'e
property for partial differential equations}, J. Math. Phys. {\bf 24} (1983) 522-526;
J. Weiss, in {\it Painlev\'e transcendents}, ed. D. Levi and P. Winternitz, Plenum
Press 1992 and references therein.}].  Yet, no Painlev\'e analysis has been done
directly in superspace.

\n {\bf OP-7:} Is it possible to formulate the Painlev\'e analysis in superspace?

Moreover, little is known concerning the solutions of super integrable systems.  The
Darboux transformation in the $N=1$ case has been worked out in
[\ref{Q.P. Liu, {\it Darboux transformations for supersymmetric Korteweg-de
Vries equations}, Lett. Math. Phys. {\bf 35} (1995) 115-122; Q.P. Liu and M. Manas
{\it Darboux transformations for SUSY integrable systems},in {\it Supersymmetry and
integrable models}, Lect. Notes in Phys. {\bf 35}, eds H. Aratyn et al, Springer
Verlag (1995), 269-281.}] but nothing has been done at this point concerning the
$N=2$ cases.\foot{Some solutions for the SKdV$_a$ equations have been reported in
[\ref{M.A. Ayari, V. Hussin and P. Winternitz, {\it Group invariant solutions for
the $N=2$ super Korteweg-de Vries equation}, J.Math. Phys. {\bf 40}
(1999) 1951-1965.}].}

Finally, unravelling the deep
relations between geometry and soliton theory has been an important theme of this
workshop;  little is known on the super version of this
connection. In particular, the super KdV equations have super Sine-Gordon relatives
and these should lead to very interesting geometrical structures.

\vskip0.3cm
\centerline{\bf Acknowledgement}
I would like to thank Z. Popowicz for his helpful comments on the recent
literature and the organisers of the workshop for their kind invitation to present this
work. 

\vskip0.3cm 

\centerline{\bf REFERENCES}
\immediate\closeout\refs \vskip 0.5cm
  \message{References}\input references
\vfill\eject

\end